\documentstyle[aps]{revtex}
\begin{document}
\draft
\preprint{CGPG-00/6-5, hep-th/0006211}
\title{\bf A new holographic entropy bound from quantum geometry }
\author{Saurya Das\footnote{email:das@gravity.phys.psu.edu}$^a$,
Romesh K Kaul\footnote{email:kaul@imsc.ernet.in}$^b$ and Parthasarathi
Majumdar\footnote{email:partha@imsc.ernet.in}$^b$  }
\address{$^a$Center for Gravitational Physics and Geometry, Penn State
University, University Park, PA 16802-6300, USA}
\address{ $^b$The Institute of Mathematical Sciences, Chennai
600 113, India.}
\maketitle
\begin{abstract}
A new entropy bound, tighter than the standard holographic bound due to
Bekenstein, is derived for spacetimes with non-rotating isolated horizons,
from the quantum geometry approach in which the horizon is described by
the boundary degrees of freedom of a three dimensional Chern Simons theory.
\end{abstract}

The Holographic Principle (HP) \cite{jw} - \cite{bek2} and the holographic
Entropy Bound (EB) have been the subject of a good deal of
attention lately.  In its original form \cite{jw}, \cite{thf},
the HP asserts that the maximum possible number of degrees of freedom within
a macroscopic bounded region of space is given by a quarter of the area (in
units of Planck area) of the boundary. This takes into account that a black hole
for which this boundary is (a spatial slice of) its horizon, has an entropy
which obeys the Bekenstein-Hawking area law and also the generalized second
law of black hole thermodynamics \cite{bek1}.  Given the relation between
the number of degrees of freedom and entropy, this translates into a
holographic EB valid generally for spacetimes with boundaries.

The basic idea underlying both these concepts is a network, at
whose vertices are variables that take only two values (`binary', `Boolean'
or `pixel'), much like a lattice with spin one-half variables at its sites.
Assuming that the spin value at each site is {\it independent} of that at
any other site (i.e., the spins are {\it randomly} distributed on the
sites), the dimensionality of the space of states of such a network is
simply $2^p$ for a network with $p$ vertices. In the limit of arbitrarily
large $p$, such a network can be taken to approximate the macroscopic
surface alluded to above, a quarter of whose area bounds the entropy
contained in it. Thus, any theory of quantum gravity in which spacetime
might acquire a discrete character at length scales of the order of 
Planck scale, is expected to conform to this counting and hence to the HP.

Let us consider now a slightly altered situation: one in which the binary
variables at the vertices of the network considered are no longer
distributed randomly, but according to some other distribution. Typically,
for example, one could distribute them {\it binomially}, assuming, without
loss of generality, a large lattice with an even number of vertices.
Consider now the number of cases for which the binary variable acquires
one of its two values, at exactly $p/2$ of the $p$ vertices. In
case of a lattice of spin 1/2 variables which can either point `up' or
`down', this corresponds to a situation of net spin zero, i.e., an equal
number of spin-ups and spin-downs. Using standard formulae of binomial
distributions, this number is
\begin{equation}
N({\frac{p} {2}} |a) = 2^p~ \left( \begin{array}{c}
                         p \\ p/2
                        \end{array} \right)~[a~(1-a)]^{p/2}
~, \label{bino} \end{equation}
where, $a$ is the probability of occurrence of a spin-up at any given vertex.
Clearly, this number is maximum when the probability of occurrence $a=1/2$;
it is given by $p! /(\frac{p}{2}!)^2$. Thus, the number of
degrees of freedom is now no longer $2^p$ but a smaller number. This
obviously leads to a lowering of the entropy. For very large $p$
corresponding to a macroscopic boundary surface, this number is
proportional to $2^p/p^{\frac12}$. The new EB can therefore be expressed as
\begin{equation}
S_{max}~=~ln \left( {\exp{S_{BH}} \over S_{BH}^{1/2} } \right)~,
\label{newb}
\end{equation}
where, $S_{BH}=A_H /4 l_P^2$ is the Bekenstein-Hawking entropy.
This is a tighter bound than that of ref.\cite{bek1} mentioned above.
The `tightening' of holographic EB is the subject of this paper. We
shall show below that, in the quantum geometry framework,
it is possible to have an even tighter bound than that depicted in eq. 
(\ref{newb}). 
 
There are of course examples of situations where the EB is violated
\cite{suss2}, \cite{bou} and must be generalized. However, generalizations
proposed so far \cite{bou} appear to be tied to fixed classical background
spacetimes, and may not hold when gravitational fluctuations are taken
into account \cite{smo}. In this note, we restrict ourselves to the older
version of the EB appropriate to stationary spacetimes, but with allowance
for the existence of radiation in the vicinity of the boundary. In this
sense, the appropriate conceptual framework is that of the Isolated
Horizon \cite{ash}.  We consider generic 3+1 dimensional isolated horizons
without rotation, on which one assumes an appropriate class of boundary
conditions. These boundary conditions require that the
gravitational action be augmented by the action of an $SU(2)$ Chern-Simons 
theory living on the isolated horizon \cite{ash}.  Boundary states of the 
Chern-Simons
theory contribute to the entropy. These states correspond to conformal
blocks of the two-dimensional Wess-Zumino model that lives on the spatial
slice of the horizon, which is a 2-sphere of area $A_H$.  The
dimensionality of the boundary Hilbert space has been calculated thus
\cite{km}-\cite{km2} by counting the number of conformal blocks of
two-dimensional $SU(2)_k$ Wess-Zumino model, for arbitrary level $k$ and
number of punctures $p$ on the 2-sphere. We shall show, from the formula
for the number of conformal blocks specialized to macroscopic black holes
characterized by large $k$ and $p$ \cite{km2}, that the restricted
situation described above, ensues, thus realizing a more stringent EB. We
may mention that similar ideas relating the quantum geometry approach to
the HP and EB have been pursued by Smolin \cite{smo}, although, as far as
we understand, the issue of tightening the Bekenstein bound has not been
addressed. 
 
We start with the formula for the number of conformal blocks of  
two-dimensional $SU(2)_k$ Wess-Zumino model that lives on the punctured
2-sphere. For a set of punctures ${\cal P}$ with spins $ \{j_1, j_2, \dots
j_p \} $ at punctures $\{ 1,2, \dots, p \}$, this number is given by
\cite{km}
\begin{equation}
N^{\cal P}~=~{2 \over {k+2}}~\sum_{r=0}^{ k/2}~{
{\prod_{l=1}^p sin \left( {{(2j_l+1)(2r+1) \pi}\over k+2} \right) }
\over
{\left[ sin \left( {(2r+1)  \pi \over k+2} \right)\right]^{p-2} }} ~.
\label{enpi}
\end{equation}
Observe now that Eq. (\ref{enpi}) can be rewritten as a multiple sum,
\begin{equation}
N^{\cal P}~=~\left ( 2 \over {k+2} \right) ~\sum_{l=1}^{k+1} sin^2
\theta_l~\sum_{m_1 =
-j_1}^{j_1} \cdots \sum_{m_p=-j_p}^{j_p} \exp \{ 2i(\sum_{n=1}^p m_n)~
\theta_l \}~,
\label{summ} \end{equation}
where, $\theta_l ~\equiv~ \pi l /(k+2)$. Expanding the $\sin^2
\theta_l$ and
interchanging the order of the summations, this becomes
\begin{equation}
N^{\cal P}~=~\sum_{m_1= -j_1}^{j_1} \cdots \sum_{m_p=-j_p}^{j_p}
\left[
~{\bar \delta}_{(\sum_{n=1}^p m_n), 0}~-~\frac12~
{\bar \delta}_{(\sum_{n=1}^p m_n),
1}~-~
\frac12 ~{\bar \delta}_{(\sum_{n=1}^p m_n), -1} ~\right ], \label{exct}
\end{equation}
where, we have used the standard resolution of the periodic Kronecker
deltas in terms of exponentials with period $k+2$,
\begin{equation}
{\bar \delta}_{(\sum_{n=1}^p m_n), m}~=~ \left( 1 \over {k+2} \right)~
\sum_{l=0}^{k+1} \exp
\{2i~[ (\sum_{n=1}^p m_n)~-~m] \theta_l \}~. \label{resol}
\end{equation}
 
Our interest focuses on the limit of large $k$ and $p$, appropriate to
macroscopic black holes of large area. Observe, first of all, that as $k
\rightarrow \infty$, the periodic Kronecker delta's in (\ref{resol})
reduce to ordinary Kronecker deltas,
\begin{equation}
\lim_{k \rightarrow \infty}~{\bar \delta}_{m_1+m_2+ \cdots +m_p,m}~=~
\delta_{m_1+m_2+ \cdots +m_p,m}~. \label{kinf}
\end{equation}
In this limit, the quantity $N^{\cal P}$ counts the number of $SU(2)$
singlet states, rather than $SU(2)_k$ singlets states. For a given set of punctures
with $SU(2)$ representations on them, this number is larger than the corresponding
number for the affine extension. This is desirable for the purpose of deducing
an upper bound on the number of degrees of freedom in any spacetime.

Next, recall that the eigenvalues of the area operator for the horizon,
lying within one Planck area of the classical horizon area $A_H$, are
given by
\begin{equation}
{\hat A}_H~\Psi_S~=~8\pi \beta
~l_{P}^2~\sum_{l=1}^p~[j_l(j_l+1)]^{\frac12}~\Psi_S~,
\label{area} \end{equation}
where, $l_{P}$ is the Planck length, $j_l$ is the spin on the $l$th
puncture on the 2-sphere and $\beta$ is the Barbero-Immirzi parameter
\cite{barb}. We consider a large fixed classical area of the horizon,
and ask what the largest value of number of punctures $p$ should be,
so as to be consistent with (\ref{area}); this is clearly obtained when
the spin at {\it each} puncture assumes its lowest nontrivial value of
1/2, so that, the relevant number of punctures $p_0$ is given by
\begin{equation}
p_0~=~{A_H \over 4 l_{P}^2}~{\beta_0 \over \beta}~, \label{pmax}
\end{equation}
where, $\beta_0=1/\pi \sqrt{3}$. We are of course interested in the case
of very large $p_0$.

Now, with the spins at all punctures set to 1/2, the number of states for
this set of punctures ${\cal P}_0$ is given by
\begin{equation}
N^{{\cal P}_0}~=~\sum_{m_1= -1/2}^{1/2} \cdots \sum_{m_{p_0}=-1/2}^{1/2}
\left[ ~\delta_{(\sum_{n=1}^{p_0} m_n), 0}~-~\frac12~
\delta_{(\sum_{n=1}^{p_0} m_n),
1}~-~\frac12 ~\delta_{(\sum_{n=1}^{p_0} m_n), -1} ~\right ] \label{excto}
\end{equation}
The summations can now be easily performed, with the result:
\begin{equation}
N^{{\cal P}_0}~=~\left( \begin{array}{c}
                         p_0 \\ p_0/2
                        \end{array} \right)
                 ~ - ~\left( \begin{array}{c}
                         p_0 \\ (p_0/2-1)
                         \end{array} \right)  ~\label{enpo}
\end{equation}
There is a simple intuitive way to understand the result embodied in
(\ref{enpo}). This formula simply counts the number of ways of making
$SU(2)$ singlets from $p_0$ spin $1/2$ representations. The first term 
corresponds to the number of states with net $J_3$ quantum number $m=0$
constructed by placing $m=\pm 1/2$ on the punctures.  However, this 
term by itself {\it overcounts} the number of SU(2) singlet states,
because even non-singlet states (with net integral spin, for $p$ is 
an even integer) have a net $m=0$ sector. Beside having a sector with 
total $m=0$, states with net integer spin have, of course, a sector with 
overall $m=\pm 1$ as well. The second term basically eliminates these
non-singlet states with $m=0$, by counting the number of states 
with net $m=\pm 1$ constructed from $m=\pm 1/2$ on the $p_0$
punctures.  The difference then is the net number of $SU(2)$ singlet 
states that one is interested in for that particular set of punctures.

To get to the entropy from the counting of the number of conformal blocks,
we need to calculate $N_{bh}=\sum_{\cal P}~N^{\cal P}$, where, the sum
is over all sets of punctures. Then, $S_{bh}~=~ln N_{bh}$. 

It may be pointed out that the first term in (\ref{enpo}) also has another
interpretation. It represents the counting of boundary states
for an effective $U(1)$ Chern-Simons theory. It counts the number of ways unit
positive and negative $U(1)$ charges can be placed on the punctures to yield a 
vanishing total charge. This would then correspond to an entropy bound given by
the same formula (\ref{newb}) above for binomial distribution of charges. 

On the other hand the combination of both terms in (\ref{enpo}), which
corresponds to counting of states in the $SU(2)$ Chern-Simons theory, yields
an even tighter bound for entropy than that in eq. (\ref{newb}).  One can 
show that \cite{ash3}, the contribution to $N_{bh}$ for this set of punctures
${\cal P}_0$ with all spins set to 1/2, is by far the dominant
contribution; contributions from other sets of punctures are far smaller in
comparison. Thus, the entropy of an isolated horizon is given by the
formula derived in ref. \cite{km2}. We may mention that very recently
Carlip \cite{car2} has presented  compelling arguments that this formula may 
possibly be of a universal character. Here, the formula follows readily from
eq. (\ref{enpo}) and Stirling approximation for factorials of large integers. 
The number of punctures $p_0$ is rewritten in terms of  area $A_H$ through
eq. (\ref{pmax}) with the identification $\beta~=~\beta_0~ln2$. This allows 
us to write the entropy of an isolated horizon in terms of a power series in 
horizon area $A_H$:
\begin{equation}
S_{bh}~=~ln N^{{\cal P}_0}~=~  {A_H\over{4l_p^2}} ~-~{3\over 2}~ln\left({A_H\over{4l_p^2}} 
\right)~-~{1\over2}~ln\left({\pi \over{8(ln2)^3}} \right) ~-~O(A_H^{-1}). \label{main}
\end{equation}
Notice that the constant term here is negative and  so is the order $A_H^{-1}$ term. 
This then implies that the entropy is bound from above by a tighter bound which
can be written in terms of Bekenstein-Hawking entropy ($S_{BH}~=~ A_H/{4l_p^2})$
as:
\begin{equation}
S_{max}~=~ln \left( {\exp{S_{BH}} \over S_{BH}^{3/2} } \right)~
\label{newb1}
\end{equation}
Inclusion of other than spin $1/2$ representations on the punctures does not
affect this bound. For example, we may place spin 1 on one or more punctures
and spin $1/2$ on the rest. The number of ways singlets can be made from 
this set of representations can be computed in a straight forward way. 
Adding these new states to the  already counted ones above, just 
changes the constant and order $A_H^{-1}$ 
terms in formula (\ref{main}). However, these additional terms continue
to be  negative and hence the entropy bound (\ref{newb1}) still
holds.\footnote{Using the Cardy formula with the prefactor (\'a la Carlip
\cite{car2}) appears \cite{jing} to lead to entropy corrections
for certain black holes not in accord with eq. (\ref{newb1}) (although the
bound (\ref{newb}) is indeed respected). This could be an artifact of the
application of the Cardy formula. We refrain from further comment on these
works since the precise relation of the Cardy formula approach to the
present framework is not clear.}
 
The steps leading to the EB now follows the standard route of deriving the
Bekenstein bound (see, e.g., \cite{smo}): we assume, for simplicity that
the spatial slice of the boundary of an asymptotically flat spacetime has
the topology of a 2-sphere on which is induced a spherically symmetric
2-metric. Let this spacetime contain an object whose entropy exceeds the
bound. Certainly, such a spacetime cannot have an isolated
horizon as a boundary, since then, its entropy would have been subject to
the bound. But, in that case, its energy should be less than that of a
black hole which has the 2-sphere as its (isolated) horizon. Let us now add
energy to the system, so that it does transform adiabatically into a black
hole with the said horizon, but without affecting the entropy of the
exterior. But we have already seen above that a black hole with such 
a horizon must respect the bound; it follows that the starting
assumption that the object, to begin with, had an entropy violating
the bound is not tenable.
 
There is, however, an important caveat in the foregoing argument. Strictly
speaking, there is as yet no derivation of the second law of black hole
mechanics within the framework of the isolated horizon. But, that is perhaps
not a conceptual roadblock as far as deriving the EB is concerned. One has
to assume that if matter or radiation crosses the isolated horizon
adiabatically in small enough amounts, the {\it isolated} character of the
horizon will not be seriously affected. This is perhaps not too drastic an
assumption. Thus, for a large class of spacetimes, one may propose 
Eq.(\ref{newb1}) as the new holographic entropy bound.

Finally, we should mention that we prefer to think of the above
holographic principle and the consequent entropy bound as `weak' rather
than `strong' in the sense of Smolin \cite{smo}. 

The work of SD was supported by NSF grant NSF-PHY-9514240 and by Eberley
Research Funds of Penn State University. PM thanks J. Ambjorn, A.
Ashtekar, A. Ghosh, H.  Nicolai, S. Kalyana Rama, R. Loll and L. Smolin for
illuminating discussions and the Center for Gravitational Physics and
Geometry at Penn State University, the Niels Bohr Institute and the Albert
Einstein Institute for their very kind hospitality during which this work
was completed.

\end{document}